# Weaknesses of a dynamic identity based authentication protocol for multi-server architecture

Weiwei Han

*School of Mathematics & Computer Science, Guangdong University of Business Studies, Guangzhou, China*

Email: hww_2006@163.com

**Abstract**: Recently, Li et al. proposed a dynamic identity based authentication protocol for multi-server architecture. They claimed their protocol is secure and can withstand various attacks. But we found some security loopholes in the protocol. Accordingly, the current paper demonstrates that Li et al.'s protocol is vulnerable to the replay attack, the password guessing attack and the masquerade attack.

## 1. Introduction

Generally, if a user wants to use numerous different network services, he must register himself to every service providing server. It is extremely hard for users to remember these different identities and passwords. In order to resolve this problem, various multi-server authentication protocols have been proposed. Recently, Li et al. [1] proposed an authentication protocol for multi-server environments and claimed their protocol could withstand various attacks. In this paper we will show Tsaur et al.'s protocol is vulnerable to the password guessing attack and a masquerade attack.

The organization of the letter is sketched as follows. The Section 2 gives a brief review of Li et al.'s protocol. The weanesses of Li et al.'s protocol are shown in Section 3. Finally, we give some conclusions in Section 4.

## 2. Li et al.'s protocol

Li et al.'s protocol involves three participants, i.e.,the user($U_i$), the service provider server ($S_j$) and the control server ($CS$). It is as summed that $CS$ is a trusted party responsible for the registration and authentication of the $U_i$ and $S_j$. $CS$ chooses the master secret key x and a secret number y. When $S_j$ register himself with $CS$ use his identity $SID_j$, the control server $CS$ computes $h(SID_j \| y)$ and $h(x \| y)$, then $CS$ shares $h(SID_j \| y)$ and $h(x \| y)$ with $S_j$. As shown in Fig. 1.[1], Li et al.'s protocol consists of three phases: Registration phase, Log-in and Authentication and session key agreement phase.

**2.1. Registration phase**

Suppose that user $U_i$ want to get service granted from the server $S_j$. $U_i$ first chooses his/her identity $ID_i$, password $P_i$ and a random number $b$, and then sends $ID_i$ and $A_i = h(b \| P_i)$ to $CS$ via a secure channel. $CS$ will perform the following steps:

1) $CS$ computes $B_i = h(ID_i \| x)$, $C_i = h(ID_i \| h(y) \| A_i)$, $D_i = B_i \oplus h(ID_i \| A_i)$ and $D_i = B_i \oplus h(y \| x)$. Then

2) $CS$ stores $C_i, D_i, E_i, h(\cdot)$ and $h(y)$ to the memory of a smart card and issue this smart card to $U_i$.

3) After receiving the smart card, $U_i$ inputs $b$ into it and finishes the registration.

**2.2. Login phase**

Once the user $U_i$ wants to login to the server $S_j$, he will perform the following login steps.

1) The user $U_i$ inputs his identity $ID_i$ and the password $P_i$ into the terminal. The smart card computes $A_i = h(b \| P_i)$ and $C'_i = h(ID_i \| h(y) \| A_i)$. Then the smart card checks whether $C_i$ and $C'_i$ are equal. If they are not equal, the smart card stops the session.

2) The smart card generates a random number $N_{i1}$, computes $B_i = D_i \oplus h(ID_i \| A_i)$, $F_i = h(y) \oplus N_{i1}$, $P_{ij} = E_i \oplus h(h(y) \| N_{i1} \| SID_j)$, $CID_i = A_i \oplus h(B_i \| F_i \| N_{i1})$ and $G_i = h(B_i \| A_i \| N_{i1})$. At last, the smart card sends $M_1 = (F_i, G_i, P_{ij}, CID_i)$ to $S_j$.

## 2.3 Authentication and session key agreement phase

1) Upon receiving the message $M_1$, $S_j$ generates a random number $N_{i2}$, computes $K_i = h(SID_j \| y) \oplus N_{i2}$ and $M_i = h(h(x \| y) \| N_{i2})$. At last, $S_j$ sends $M_2 = (F_i, G_i, P_{ij}, CID_i, SID_j, K_i, M_i)$ to $CS$.

2) After receiving $M_2$, $CS$ computes $N_{i2} = h(SID_j \| y) \oplus K_i$ and checks whether $M_i$ and $h(h(x \| y) \| N_{i2})$ are equal. If they are not equal, $CS$ stops the session.

3) $CS$ computes $N_{i1} = h(y) \oplus F_i$, $B_i = P_{ij} \oplus h(h(y) \| N_{i1} \| SID_j) \oplus h(y \| x)$, and $A_i = CID_i \oplus h(B_i \| F_i \| N_{i1})$. Then $CS$ checks whether $G_i$ and $h(B_i \| A_i \| N_{i1})$ are equal. If they are not equal, $CS$ stops the session.

4) $CS$ generates a random number $N_{i3}$, computes $Q_i = N_{i1} \oplus N_{i3} \oplus h(SID_j \| N_{i2})$, $R_i = h(A_i \| B_i) \oplus h(N_{i1} \oplus N_{i2} \oplus N_{i3})$, $V_i = h(h(A_i \| B_i) \| h(N_{i1} \oplus N_{i2} \oplus N_{i3}))$, $T_i = N_{i2} \oplus N_{i3} \oplus h(A_i \| B_i \| N_{i1})$. Then $CS$ sends $M_3 = (Q_i, R_i, V_i, T_i)$ to $S_j$.

5) Upon receiving the message $M_3$, $S_j$ computes $N_{i1} \oplus N_{i3} = Q_i \oplus h(SID_j \| N_{i2})$ and $h(A_i \| B_i) = R_i \oplus h(N_{i1} \oplus N_{i2} \oplus N_{i3})$. Then $S_j$ checks whether $V_i$ and $h(h(A_i \| B_i) \| h(N_{i1} \oplus N_{i2} \oplus N_{i3}))$ are equal. If they are not equal, $S_j$ stops the session. At last, $S_j$ sends $M_4 = (V_i, T_i)$ to $U_i$.

6) Upon receiving the message $M_4$, $U_i$'s smart card computes $N_{i2} \oplus N_{i3} = T_i \oplus h(A_i \| B_i \| N_{i1})$ and checks whether $V_i$ and $h(h(A_i \| B_i) \| h(N_{i1} \oplus N_{i2} \oplus N_{i3}))$ are equal. If they are not equal, $U_i$ stops the session.

Finally, the user $U_i$, the server $S_j$ and the control server $CS$ agree on a common session key as $SK = h(h(A_i \| B_i) \| N_{i1} \oplus N_{i2} \oplus N_{i3})$.

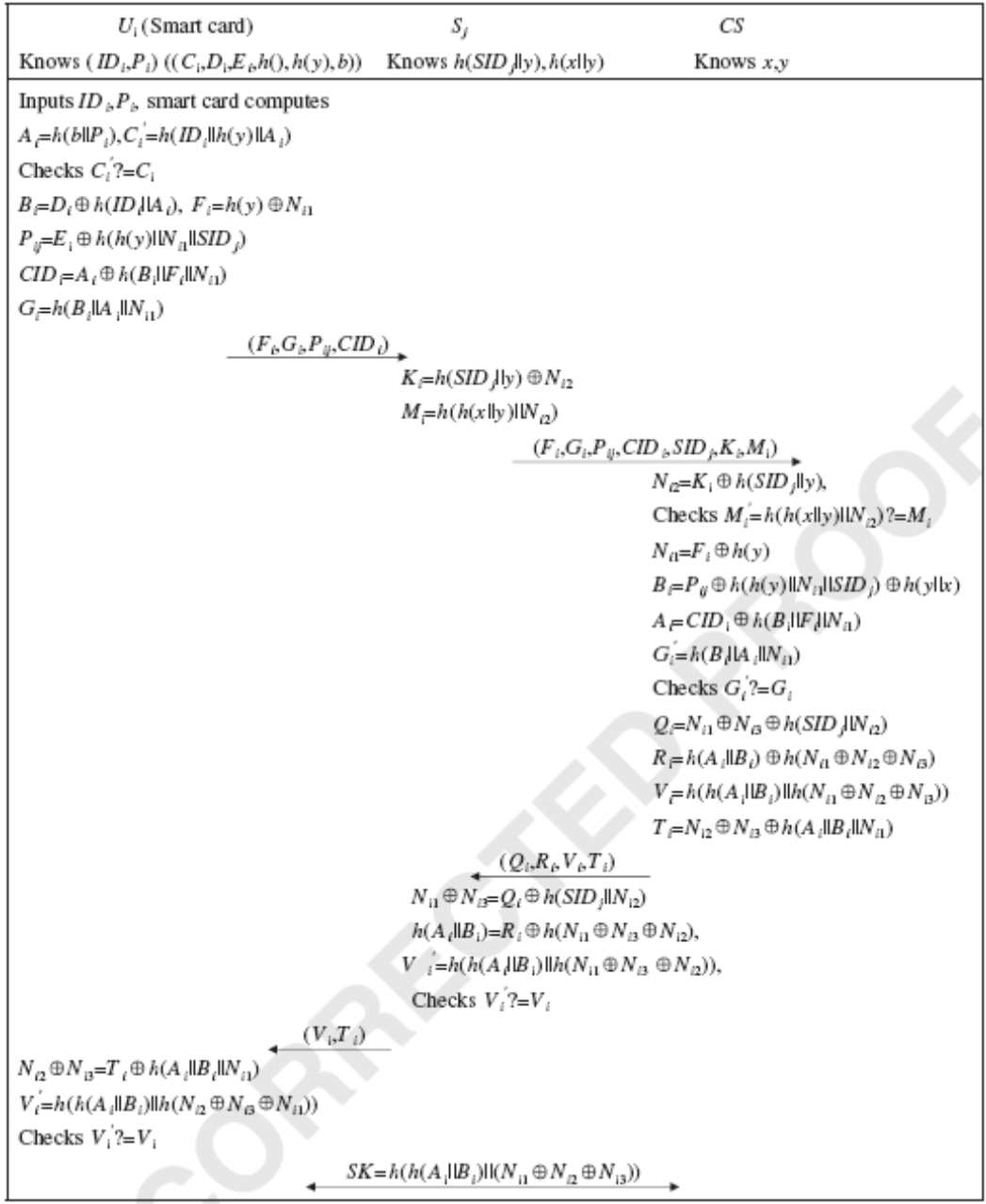

Fig. 1. Work flow of Li et al.'s protocol

## 3. Weaknesses of Li et al.'s protocol

### 3.1. Password Guessing Attack

Kocher et al. [2] and Messerges et al. [3] have pointed out that all existent smart cards are vulnerable in that the confidential information stored in the device could be extracted by physically monitoring its power consumption; once a card is lost, all secrets in it may be revealed. To evaluate the security of smart card based user authentication, we assume the capabilities that an adversary $\mathscr{A}$ may have as follows:

1) The adversary has total control over the communication channel between the users and the server in the login and authentication phases. That is, $\mathscr{A}$ may intercept, insert, delete, or modify any message in the channel.

2) $\mathscr{A}$ may (i) either steal a user's smart card and then extract the information from it, (ii) or obtain a user's password, (iii) but not both (i) and (ii).

Suppose an adversary $\mathscr{A}$ has stolen $U_i$'s smart card and extracted the stored values $C_i, D_i, E_i, h(\cdot)$ and $h(y)$, where $A_i = h(b \| P_i)$, $B_i = h(ID_i \| x)$, $C_i = h(ID_i \| h(y) \| A_i)$, $D_i = B_i \oplus h(ID_i \| A_i)$ and $D_i = B_i \oplus h(y \| x)$. Then the attacker $\mathscr{A}$ could find the password guessing by performing the following procedure.

1) $\mathscr{A}$ guesses a password $P_i'$ and a identity $ID_i'$.

2) $\mathscr{A}$ computes $A_i' = h(b \| P_i')$ and $C_i' = h(ID_i' \| h(y) \| A_i)$.

3) $\mathscr{A}$ checks whether $C_i'$ and $C_i$ are equal. If they are equal, $\mathscr{A}$ finds the correct password. Otherwise, $\mathscr{A}$ repeats 1)-3) until finding the correct password.

From the above description, we know the adversary can get the password. Therefore, Li et al.'s protocol is vulnerable to the password guessing attack.

### 3.2 Masquerade Attack

Let $U_t$ is a malicious user. Then he will get his secure key $C_t, D_t, E_t, h(\cdot)$ and $h(y)$, where $A_t = h(b \| P_t)$, $B_t = h(ID_t \| x)$, $C_t = h(ID_t \| h(y) \| A_t)$, $D_t = B_t \oplus h(ID_t \| A_t)$ and $D_t = B_t \oplus h(y \| x)$. Then $U_t$ could impersonate the user $U_i$ through the following steps.

$U_t$ generates a random number $N_{i1}$, computes $B_t = D_t \oplus h(ID_t \| A_t)$, $F_i = h(y) \oplus N_{i1}$, $P_{ij} = E_t \oplus h(h(y) \| N_{i1} \| SID_j)$, $CID_i = A_t \oplus h(B_t \| F_i \| N_{i1})$ and $G_i = h(B_t \| A_t \| N_{i1})$. At last, the smart card sends $M_1 = (F_i, G_i, P_{ij}, CID_i)$ to $S_j$.

It is easy to verify that $M_1 = (F_i, G_i, P_{ij}, CID_i)$ could pass $CS$'s verification. Then $U_t$ could impersonate $U_i$ successfully.

From the above description, we know $U_t$ could impersonate $U_i$ successfully. Therefore, Li et al.'s protocol is vulnerable to the masquerade attack.

### 3.3 Replay Attack

The adversary $\mathscr{A}$ has total control over the communication channel between the users and the server in the login and authentication phases. That is, $\mathscr{A}$ may intercept, insert, delete, or modify any message in the channel.

$\mathscr{A}$ could intercept a legal message $M_1 = (F_i, G_i, P_{ij}, CID_i)$ generated by a user $U_i$. Then $\mathscr{A}$ could send it to the server $S_j$. Since $M_1$ is generated by $U_i$, then it could pass the verification of $CS$ and $S_j$.

From the above description, we know $U_t$ could impersonate $U_i$ successfully. Therefore, Li et al.'s protocol is vulnerable to the replay attack.

## 4. Conclusion

Recently, Li et al. proposed an authentication protocol for multi-server environments and demonstrated its immunity against various attacks. However, after review of their protocol and analysis of its security, three kinds of weaknesses are presented in different scenarios. The analyses show that the protocol is insecure for practical application.